\documentstyle[multicol,aps]{revtex}
\begin{document}
%\date{}
\title{Kosterlitz-Thouless and Manning Condensation}
\author{Yan Levin \\
Instituto de F\'{\i}sica, Universidade Federal
do Rio Grande do Sul\\ Caixa Postal 15051, CEP 91501-970
, Porto Alegre, RS, Brazil\\
{\small levin@if.ufrgs.br}}

\maketitle
\begin{abstract}
A comparison between the Kosterlitz-Thouless theory of metal insulator
transition in a two dimensional plasma and a counterion condensation in a
polyelectrolyte solution is made. It is demonstrated that, unlike some of
the recent suggestions, the counterion condensation and the Kosterlitz-Thouless
transition are distinct.    
\end{abstract}
\begin{multicols}{2}

\section{\bf Introduction}

        The polyelectrolyte solutions present one of the outstanding
problems in the field of physical chemistry. Although the work of Debye
and H\"uckel \cite{DE23} shed some light on the unusual behavior of simple
symmetric electrolytes, the properties of the polyelectrolyte solutions
still remain to large extent ununderstood. In general a polyion can be a
flexible polymer chain or a rigid molecule of specific shape some of whose
monomers are ionized. For now, we shall confine our discussion to rigid
polyelectrolytes whose polyions can be modeled as cylinders or spheres, since
as the reader will see, even these significantly simpler systems already
present more than enough complications. In particular even such basic
question, as what is the form of electrostatic interaction between two
polyions, still remains controversial. A prototypical molecule whose shape
can be modeled as a cylinder is a DNA. It should be noted, however, that
even in this case the approximation of replacing a long molecule by an infinite
cylinder will only be valid if the persistence lenght is sufficiently
greater than the Debye screening lenght inside the solution. As an example of
spherical polyions, one can consider various colloidal suspensions, for
example the ones made of latex or polystyrene particles.

        In this note we shall confine our attention to the case of rigid
cylindrical polyions. In particular, our discussion is motivated by the
recent observation of Kholodenko and Beyerlein \cite{KH95} that the commonly
observed counterion condensation during which a finite fraction of
 disassociated
counterions recondenses onto polyions, is a ``special case of the
Kosterlitz-Thouless \cite{KO73} (KT)-like phase transition'' \cite{KH95}.
Here we shall demonstrate that the KT transition and the Manning's counterion
condensation \cite{MA69} have nothing in common. The erroneous conclusion reached
by the above authors can be attributed to the confusion over the system of units
used in the original KT theory.

        To begin our discussion let us first define the Primitive Model of
Polyelectrolyte (PMP) \cite{LE96}. It consists of  $N$ cylindrical polyions
of a cross-sectional diameter $a_p$ and lenght $L$, each carrying 
$P$ ionized groups
of charge $-q$ spaced uniformly, with a separation $b$, along the axis of a
cylinder. A total of $PN$ counterions are 
present to preserve the overall charge
neutrality of the system. The counterions are assumed to be spherical in shape
with a diameter $a_c$, each carring a charge $q$. The whole system is
confined inside a region of volume  $V$. It will prove convenient to
define the distance of closest approach between the centers of a polyion and
a counterion, $a \, = \, (a_p \, + \, a_c)/2$. The solvent is modeled as a uniform
medium of a dielectric constant $D$. The bare 
interaction between a sufficiently
long polyion and a counterion can then  be approximated by
\begin{equation}
\label{eq1}
\phi (r) \; \; = \; \; \left\{
\begin{array}{cl} +
\infty \, , & r\, < a, \\
-2q(\sigma/D) \ln (r/a), & r\geq a,
\end{array}
\right.
\end{equation}
where $\sigma \, = \, -Pq/L \, = \, -q/b$. If we make the replacement
$\sigma \, \rightarrow \, -q$ (or $b \, \rightarrow \, 1$), the
logarithmic potential which appears in the above expression is
exactly the same as for two oppositely charged hard disks in two
dimensions (assuming we keep the same Poisson equation, $\nabla^2 \phi \,
= \, -4 \pi \rho_q /D$, as in three dimensions). Furthermore, it is well
known that when the temperature of a two dimensional plasma is lowered
below a certain value, all the ions associate into dipolar pairs, changing
the properties of plasma from being a conductor to an insulator \cite{KO73}.
This is a continuous thermodynamic transition, which happens to be of infinite
order. Is it possible that the counterion condensation observed in
polyelectrolytes is a realization of this transition? In order to answer
this question we shall present two simple mean-field theories, one for a
two dimensional Coulomb gas and another for a polyelectrolyte solution
in three dimensions.

\section{\bf 2d Coulomb Gas}

        Our system will consist of hard disks of diameter $a$, half of which
carry the charge $+q$, while the other half carry the charge $-q$F.
 As is usual for the restricted primitive model (RPM),
the solvent is modeled as a uniform medium of a dielectric constant $D$. Due
to strong electrostatic interactions we expect that some of the ions will
associate forming dipolar pairs \cite{LE94}. The total density of hard
spheres is $\rho \, = \, \rho_+ + \rho_- + 2 \rho_2$, where $\rho_+ \, =
\, \rho_- \equiv \rho_1 /2$ is the density of free unassociated ions,
while the $\rho_2$ is the density of dipolar pairs. In order to explore
the thermodynamic properties of this plasma, we need the free energy
for the system. We shall construct this free energy out of two parts:
the entropic free energy associated with the momentum degrees of freedom
in the partition function, and the electrostatic free energy due to the
interactions between the ions. It is a simple matter to write the expression
for the entropic part of the free energy, since it corresponds to that of an
ideal gas. Defining the Helmholtz free energy density as $f \, = \, -F/V$
we have
\begin{equation}
\label{eq2}
\beta f^{ent} \, = \, \sum_s \, \rho_s \, [1 \, - \, \ln(\rho_s
\Lambda_s^{d \mid s \mid} / \zeta_s)],
\end{equation}
where $d$ is the dimensionality, $\beta \, = \, 1/k_BT, \; \Lambda_s$ is the
de Broglie thermal wavelength, $\zeta_s$ is the internal partition function for
species $s \, = \, +,-,2,$ and $\mid s \mid$ is how many particles constitute
a specie, i.e. $\mid + \mid \, = \, \mid - \mid \, = \, 1, \mid 2 \mid \, = \, 2$. The electrostatic free energy is easily calculated in the spirit of the
original Debye-H\"uckel theory \cite{LE94}. We find
\begin{equation}
\label{eq3}
\beta f^{el} \, = \, \frac{1}{2 \pi a^2} \, \ln[\kappa aK_1 (\kappa a)],
\end{equation}
where $K_1$ is the modified Bessel function of first order,
 $(\kappa a)^2 \, = \, 4 \pi \rho_1^*/T^*$, and the reduced density and
temperature are respectively $\rho^* \, = \, \rho a^2$ and $T^* \, = \, k_B
T/q^2$. The densities of the free and the associated ions are related through
the law of mass action $\mu_+ \, = \, \mu_- \, = \, \mu_2$, where the
chemical potentials are determined from the free energy $\mu_s \, = \,
-\partial f / \partial \rho_s$. We find
\begin{equation}
\label{eq4}
\rho_2 \, = \, \frac{1}{4} K(T)\rho_1^2 e^{2 \mu^{ex}},
\end{equation}
where the excess chemical potential is $\mu^{ex} \, = \, -\partial
f^{el}/\partial \rho_1$ and the equilibrium constant $K(T) \, = \,
\zeta_2(T)/\zeta_+ \zeta_-$. For the purpose of the present exposition it is
sufficient to know that the equilibrium constant remains finite in the
limit $\rho_1 \rightarrow 0$, for further discussion we refer the interested
reader to \cite{LE94}. In the limit of small densities the excess chemical
potential can be expanded in powers of $\rho_1, \mu^{ex} \, = \,
-[\gamma_E + \ln (\kappa a/2)]/T^* +O(\rho_1)$. Substituting this expression
into Eq. \ref{eq4}, we observe that $\rho_2^* \, \approx \, \rho_1^{*2}
e^{- \ln(\rho_1^*)/T^*} \, = \, \rho_1^{*(2-1/T^*)}$ \cite{LE94}, \cite{HO80}.
In particular, we see that for a {\it fixed} density $\rho$, as the
temperature is reduced to $T^* \rightarrow T_{KT}^* \, = \, 1/2$ the density of
free ions, $\rho_1 \rightarrow 0$. This is exactly the Kosterlitz-Thouless
metal insulator transition. Since the Debye screening length is inversely
proportional to $\sqrt{\rho_1}$, we see that it will diverge as $T \rightarrow
T_{KT}^+$.

        Let us now take a look at what happens in the case of cylindrical
polyelectrolytes \cite{LE96}.

\section{Polyelectrolyte solution}

        We shall work in the context of the PMP defined above. Just as in
the case of simple electrolyte addressed in the previous section, we
expect that the strong electrostatic interaction between the polyions and
the counterions will result in formation of clusters composed of one polyion
and $1 \leq n \leq Z$ condensed 
counterions. We are then lead to two conservation laws:
\begin{equation}
\label{eq5}
\rho \, = \, \sum_{n=0}^{P} \rho_n
\end{equation}
\begin{equation}
\label{eq6}
P \rho \, = \, \rho_+ + \sum_{n=0}^{P} n \rho_n ,
\end{equation}
where $\rho$ is the total density of the polyions,$\rho_0$
is the density of free polyions, $\rho_n$ is the
density of the clusters of size $n$, and $\rho_+$  is the density of
free, unassociated, counterions. In order to explore the thermodynamic
properties of the PMP we require the free energy. As in the case of
simple electrolyte, this can be constructed as a sum of the entropic
and the electrostatic contributions. The entropic part of the free
energy is given by the Eq. \ref{eq2} with $s \, = \, \{+,0 \leq n \leq Z \}$.
The electrostatic contribution, however, is significantly more complex
than for simple electrolyte. It consists of the free energies due to the
polyion-counterion, polyion-polyion, and the counterion-counterion
interactions. In general it is quite difficult to take a full account
of all these effects. The condensation phenomena that we are interested in
studing, however, already occurs in the limit of vanishingly small
densities, $\rho \rightarrow 0$. In this limit, it is possible to show
that the polyion-polyion and the counterion-counterion interactions
are small and the main contribution to the electrostatic free energy
comes from the polyion-counterion interaction \cite{LE96}. This
contribution can be calculated in the spirit of the Debye-H\"uckel
theory and we find \cite{LE96}
\begin{equation}
\label{eq7}
\beta f^{el} \, = \, \sum_{n=0}^P \, \frac{\rho_n^*(P-n)^2}{La^2 T^*}
\left[ \ln \left(\frac{\kappa a}{2}\right) + O (\rho_+^0) \right]
\end{equation}
where $(\kappa a)^2 \, = \, 4 \pi \rho_+^*/T^*$ and the reduced density and
temperature are now $\rho^* \, = \, \rho a^3$ and $T^* \, = \, k_BTDa/q^2$
respectively. The distribution of cluster sizes can be found from the law of
mass-action, $\mu_n \, = \, \mu_0 + n\mu_+$, which reduces to
\begin{equation}
\label{eq8}
\rho_n^* \, = \, K_n(T)\rho_0^*(\rho_+^*)^n\mbox{e}^{\beta \mu_0^{ex}
+n\beta \mu_+^{ex} - \beta \mu_n^{ex}}
\end{equation}
where the excess chemical potentials, $\mu_s^{ex} \, = \, - \partial
f^{el}/\partial \rho_s$, are \newline $\beta \mu_n^{ex} \, = \,
- \frac{(P-n)^2a}{PT^*b} 
\left[ \ln (\frac{\kappa a}{2}) + O(\rho_+^0)
\right]$ and $\beta \mu_+^{ex} \, = \, O(\rho_+^0)$. The equilibrium
constant, $K(T) \, = \, \zeta_s (T)/ \zeta_+^n$, as in the case of simple
electrolyte will remain finite in the limit $\rho \rightarrow 0$.
Substituting the expressions for the excess chemical potential into Eq.
\ref{eq8} we find that $\rho_n^* \, \approx \, \rho_0^* \rho_+^{*^{g(n)}}$,
where $g(n) \, = \, n -na/T^*b+n^2a/2PT^*b$. In the limit $\rho
\rightarrow 0$ the only possible cluster has the size $n^*$, for which
the function $g(n)$ attains its minimum. In particular we see that for
$T^*<T_M^* \, = \, a/b$ the  minimum is attained when $n^* \, = \, P(1-T^*
b/a)$ counterions are associated with a polyion, while for $T^* > T_M^*$ no
clusters form and the minimum is at $n^* \, = \, 0$. In the limit
$\rho \rightarrow 0$, the formation of clusters is a continuous transition
which occurs when the temperature is lowered below $T^* \, = \, T_M^* \,
= \, a/b$.

\section{Conclusion}

        As was discussed following Eq. \ref{eq1} the isomorphism between 2d
Coulomb law and the interaction potential between a polyion and a counterion
is valid if $b \rightarrow 1 \, \, 
(\sigma \rightarrow -q)$. In these units $T_M \,
= \, q^2/k_BD$, while the Kosterlitz-Thouless transition occurs at
$T_{KT} \, = \, q^2/2k_BD$, which is half the value of the equivalent Manning
temperature. Furthermore, we would like to stress that while at KT transition
{\it all} the ions associate into dipolar pairs, thus leading to a
divergent Debye screening length, nothing like this happens in the case of
polyelectrolytes. Quite on the contrary, all the way down to zero
temperature there remain some free counterions, producing a finite screening
lenght. Finally, while the KT transition is found at non-zero density, the
sharpness (discontinuity in slope as a function of temperature) of the counterion
condensation transition will disappear with an increase of density \cite{KU}.

\section*{Acknowledgments}

At various stages of this work the author has benefited greatly from
interactions and collaborations with M.E. Fisher, X.-J. Li, M.C. Barbosa
and M.N.Tamashiro.

\end{multicols}
\end{document}